# Conductance quantization and the $0.7 \times 2e^2/h$ conductance anomaly in one-dimensional hole systems

R. Danneau[a)], W.R. Clarke, O. Klochan, A.P. Micolich, A.R. Hamilton and M.Y. Simmons

*School of Physics, University of New South Wales, Sydney NSW 2052, Australia.*

M. Pepper and D.A. Ritchie

*Cavendish Laboratory, University of Cambridge, Cambridge CB3 0HE, U.K.*

We have studied ballistic transport in a 1D channel formed using surface gate techniques on a back-gated, high-mobility, bilayer 2D hole system. At millikelvin temperatures, robust conductance quantization is observed in the quantum wire formed in the top layer of the bilayer system, without the gate instabilities that have hampered previous studies of 1D hole systems. Using source drain bias spectroscopy, we have measured the 1D subband spacings, which are 5-10 times smaller than in comparable GaAs electron systems, but 2-3 times larger than in previous studies of 1D holes. We also report the first observation of the anomalous conductance plateau at $G = 0.7 \times 2e^2/h$ in a 1D hole system.

---

[a)]Corresponding author. E-mail: r.danneau@unsw.edu.au

The conductance $G$ through a ballistic one-dimensional (1D) channel is quantized in units of $G_0 = 2e^2/h$, due to the discrete energy subbands produced by 1D confinement.[1] The resulting staircase of conductance plateaus was first observed by using surface gate techniques to define a 1D quantum wire (QW) in the two-dimensional electron gas of an $Al_xGa_{1-x}As$-GaAs heterostructure.[2] This cornerstone of mesoscopic physics has been extensively studied in electron systems, with considerable recent attention focused on an anomalous conductance plateau that occurs at $G = 0.7 \times G_0$, commonly knows as "0.7 structure" and believed to be of many-body origin.[3-9]

The study of 1D systems based on holes promise significant new physics due to two key properties. Firstly, the stronger spin-orbit coupling in holes[10] may allow their application in spintronics.[11] Secondly, the larger effective mass $m^*$ of holes compared to electrons will allow the study of 1D many-body effects such as the 0.7 structure. However, despite their potential, 1D hole systems have received only limited attention to date, mainly due to significant difficulties in achieving clear and stable conductance quantization. Previous studies report only unclear[12] or weak[13] plateaus, or significant problems with gate instabilities.[14] This is partially due to difficulties in fabricating hole samples compared to the more established electron samples. However, despite promising new physics, to some extent the larger hole mass introduces additional problems. Firstly, the larger $m^*$ reduces the 1D subband spacing (since $\Delta E_{N,N+1} \propto 1/m^*$), making it more difficult to resolve the conductance plateaus at a given temperature $T$. Secondly, the larger $m^*$ results in a mean free path several times smaller than that found in electron systems, making it more difficult to observe ballistic transport and conductance quantization, and hence requiring the use of significantly higher quality wafer.

In this letter, we report clear and robust conductance quantization in a 1D hole system defined using surface-gate techniques on a bilayer 2D hole system. These devices are sufficiently stable that in addition to the integer quantized plateaus, we also make the first observation of the anomalous plateau at $0.7 \times G_0$ in holes. Finally, we use source drain bias (SDB) measurements to determine the 1D subband spacings $\Delta E_{N,N+1}$ for the first five subbands. The values we obtain are higher than previously measured in 1D hole systems,[13] suggesting that our device design using gates above and below the QW provides better confinement of the 1D holes than in previous studies.[12-14]

The devices were fabricated from wafers grown using molecular beam epitaxy and consisting of a modulation-doped bilayer structure formed by two 20 nm GaAs quantum wells separated by a 30 nm layer of $Al_{0.3}Ga_{0.7}As$ on a conducting (311)A n+-GaAs substrate, which serves as the back-gate.[15] The thick AlGaAs barrier and the large hole mass $m^*$ limit tunneling between the two quantum wells. The Hall bar is defined using optical lithography and wet etching, and aligned along the higher-mobility [$\bar{2}$33] direction. The hole densities and mobilities of the top and the bottom layer are, respectively, $n_{s\text{-TOP}} = 1.2 \times 10^{15}$ m$^{-2}$ and $\mu_{TOP} = 92$ m$^2$V$^{-1}$s$^{-1}$, and $n_{s\text{-BOTTOM}} = 1.0 \times 10^{15}$ m$^{-2}$ and $\mu_{BOTTOM} = 87$ m$^2$V$^{-1}$s$^{-1}$. Ti/Au gates, defined using optical and electron-beam lithography, are patterned on the surface, 280 nm above the top layer. The n$^+$-GaAs back-gate is ≈ 3 μm below the bottom layer. The samples were measured in a Kelvinox 100 dilution refrigerator with a base temperature $T$ = 20 mK. Electrical measurements were performed using standard low-frequency a.c. lock-in techniques with an excitation of 20-100 μV at 17 Hz. A scanning electron micrograph and a schematic view of the device architecture are shown in Figs. 1(a) and 1(b), respectively. The two side gates (SG) are used to define the wires. The back-gate (BG) and a mid-line gate (MG) allow us to control the hole density and 1D confinement potential in the two layers, and act to stabilize the 1D system. Finally, a depletion gate (DG) allows us to pinch off the top layer and measure the bottom layer independently, while at sufficiently high bias the back-gate allows us to deplete the bottom layer and measure the top layer independently.

To illustrate the basic operation of the device, in Fig. 1(c) we show measurements of the conductance $G$ as a function of side gate voltage $V_{SG}$ at $T$ = 4.2 K, a temperature sufficiently high that quantized conductance plateaus are not observed. The top trace in Fig. 1(c) shows a four-stage depletion obtained when both layers are measured in parallel. As $V_{SG}$ is increased, we observe the following stages: (1) definition of the wire in the top layer; (2) definition of the wire in the bottom layer; (3) pinch-off of the wire in the bottom layer; and (4) pinch off of the wire in the top layer. The pinch-off of the bottom wire before the top wire is counterintuitive, but confirmed by separately measuring the conductance of the layer. This effect has already been reported in bilayer electron systems.[16,17] The lower trace in Fig. 1(c) shows the conductance of the bottom layer only, obtained with the DG biased to cut off the top layer so that all current is forced to flow in the bottom layer. In this configuration we don't see the definition of the wire in the top layer, instead seeing only the definition and pinch-off of the wire in the bottom layer.

To look for evidence of quantized conductance, we reduced the temperature to $T = 20$ mK, where the thermal smearing $kT$ is smaller than the subband spacing $\Delta E_{N,N+1}$ and repeated the measurements for a variety of different BG and MG biases. Under appropriate biasing conditions (1 V $< V_{BG} <$ 3 V and $-0.1$ V $< V_{MG} < -0.7$ V) we observe stable quantized conductance plateaus in the top wire once the bottom wire is depleted (i.e. in Stage 4 of the definition/depletion sequence shown in Fig. 1). Figure 2 shows the conductance of the top wire after a constant series resistance of 2.5 k$\Omega$, due to the ohmic contacts and adjoining 2D hole system, has been subtracted from the data. The inset to Fig. 2 shows the same four-stage definition/depletion structure as the top trace of Fig. 1. Remarkably clear quantized conductance plateaus can be observed, as well as a well-pronounced feature at $G = 0.7 \times 2e^2/h$. The high quality of the conductance plateaus lends confidence that this is the same feature as seen in 1D electron systems, and is not an artifact due to disorder or imperfect conductance quantization. Furthermore we find that this feature is robust to thermal cycling. We have also observed conductance quantization in the bottom wire, although the quality of the data (not shown) is not quite as good as in the top wire. We also note that the combination of $V_{BG}$ and $V_{MG}$ is essential to observing these plateaus.

To further demonstrate the stability of our QW, essential for future studies of the 0.7 structure in holes, we present measurements of the 1D subband energy spacing obtained using SDB spectroscopy.[18,19] In Fig. 3, we map the transconductance d$G$/d$V_{SG}$ (intensity axis) corrected by a series resistance of 2.5 kΩ as a function of $V_{SG}$ ($y$-axis) and the d.c. bias $V_{SD}$ ($x$-axis) applied across the top wire. Moving vertically upwards at $V_{SD}$ = 0 V (center of Fig. 3) light regions mark the integer-quantized conductance plateaus (small d$G$/d$V_{SG}$), and dark regions mark the conductance steps (larger d$G$/d$V_{SG}$). The applied positive (negative) $V_{SD}$ raises (lowers) the Fermi energy of the source with respect to the drain, and when the energy difference between source and drain is equal to the energy difference between the $N$ and $N+1$ subbands, an additional plateau quantized in odd-half-integer multiple of $e^2/h$ is observed, at $G = (2N + 1)/2 \times G_0$. The subband spacing $\Delta E_{N,N+1}$ is measured from the $V_{SD}$ at which these odd-half-integer plateaus are centered in Fig. 3. For example, the plateau at $3 \times e^2/h$ is centered at $V_{SD}$ = 0.41 V, giving $\Delta E_{1,2}$ ≈ 410 μeV. We find that $\Delta E_{N,N+1}$ decreases monotonically for higher $N$, so that the subband spacings go: $\Delta E_{1,2}$ ≈ 410 μeV, $\Delta E_{2,3}$ ≈ 340 μeV, $\Delta E_{3,4}$ ≈ 295 μeV, $\Delta E_{4,5}$ ≈ 280 μeV, and $\Delta E_{5,6}$ ≈ 275 μeV. To put these values into perspective, typical subband spacings for electrons in surface-gated GaAs QW range from 1.45 to 3.18 meV for the first five subbands.[20] The smaller subband spacings measured in our hole QW are consistent with the larger hole mass, since $\Delta E_{N,N+1} \propto 1/m^*$ and $m^*_{holes} \approx 5\, m^*_{electrons}$. However a direct comparison is difficult since the subband spacings are highly dependent on the heterostructure design, gate geometry and gate biases used in the experiments. Indeed the subband spacings measured here are significantly larger than the only previous report of subband spacings in 1D hole systems (90-270 μeV).[13] This may explain why our conductance plateaus much are better defined than the data in previous experiments,[12-14] and why the 0.7 structure is observed. We believe that the combination of multiple gates, in addition to improving the stability of the device, provide the stronger 1D confinement required to enhance $\Delta E_{N,N+1}$ and reduce the effect of thermal smearing.

In conclusion, we have studied the ballistic transport of holes through a 1D quantum point contact structure formed in the top layer of an $Al_xGa_{1-x}As$-GaAs bilayer hole device, with three key findings. Firstly, we observe stable conductance quantization with clearly defined plateaus, in contrast to previous studies, which report weak plateau structures and significant issues with gate instabilities. Secondly, the improved stability of our device allowed us to make the first observation of the anomalous plateau at $0.7 \times G_0$ in a 1D hole system. Finally, we measure the 1D subband spacing in our device and find values more than double those previously reported in 1D holes.[13] Our results demonstrate that it is possible to fabricate and measure stable low-dimensional hole systems and make it possible to explore the behavior of the 0.7 structure. Furthermore, our work opens up the possibility of using ballistic quantum wires for spin-manipulation and spin-filtering applications.[11]

We thank U. Zülicke and S. Y. Cho for helpful discussions and acknowledge support from the Australian Research Council and EPSRC.

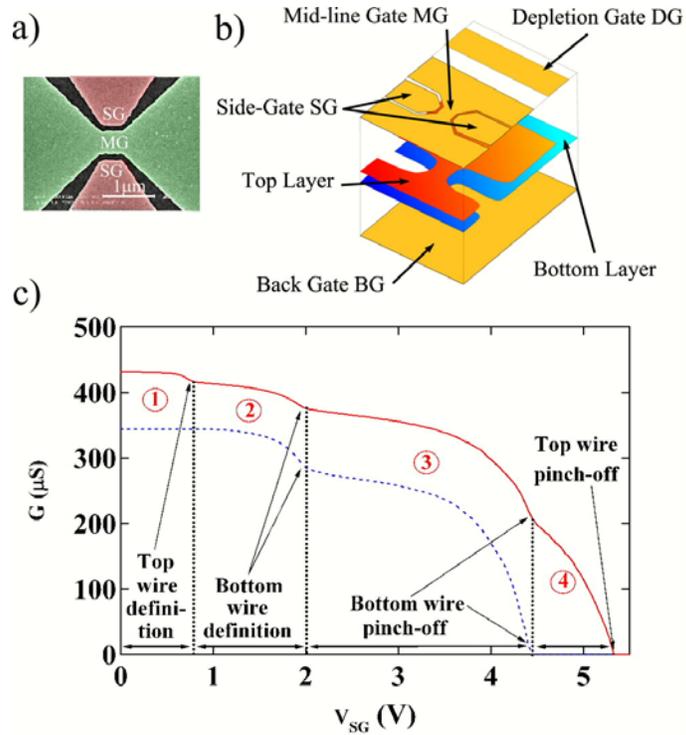

Figure 1: a) Scanning electron micrograph of the device showing the two 400 nm long side-gates (SG) separated by a 600 nm gap and the 400 nm wide mid-line gate (MG). b) Schematic view of the device. c) A plot of the conductance $G$ though the device vs voltage applied to the side-gates $V_{SG}$ at temperature $T = 4.2$ K. In the top trace, current flows through both layers, and the four stage definition/depletion discussed in the text. The bottom trace shows the same measurement when current only flows in the lower layer (by pinching off contact to the upper layer using the DG).

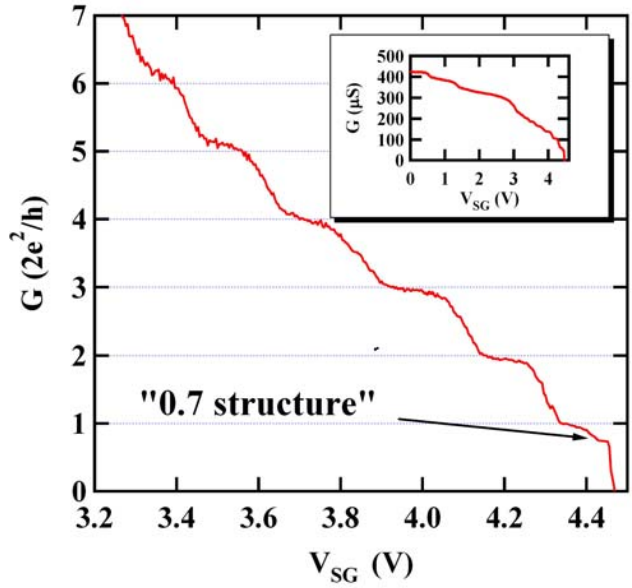

Figure 2: $G$ vs $V_{SG}$ at $T = 20$ mK, showing the quantized conductance of the top wire. The arrow highlights the anomalous plateau at $0.7 \times 2e^2/h$ in a 1D hole system. Data have been corrected for a series resistance of 2.5 k$\Omega$, and were taken with $V_{BG} = 2.5$ V and $V_{MG} = -0.7$ V. The inset shows the raw data, which exhibits the same four-stage definition/depletion shown in Fig. 1, with the addition of quantized conductance plateaus in the top wire.

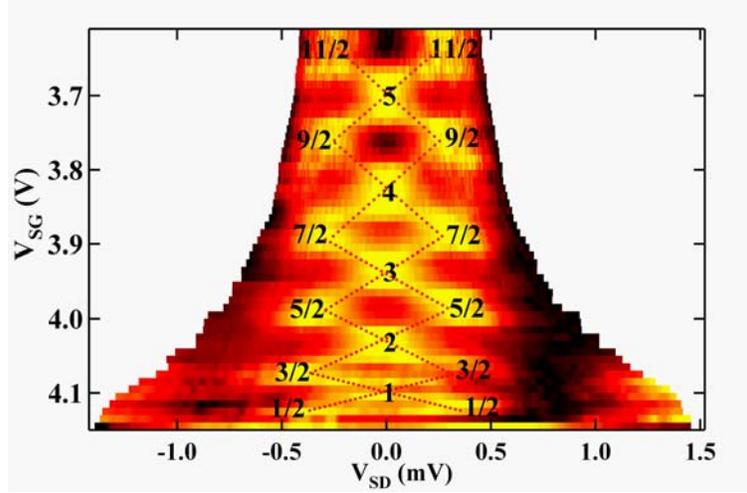

Figure 3: Colormap of the transconductance $dG/dV_{SD}$ (intensity) as a function of $V_{SG}$ (y-axis) and the applied d.c. source-drain bias $V_{SD}$ (x-axis) obtained at $T = 20$ mK, $V_{BG} = 2.5$ V and $V_{MG} = -0.5$ V. The data have been corrected for the voltage drop across the series resistance. The plateaus (small $dG/dV_{SD}$) show up as the light regions, with the steps between them (larger $dG/dV_{SD}$) appearing dark. The dashed lines serve as a guide to the eye, highlighting the standard plateaus and the additional odd-integer plateaus used to determine the subband spacings $\Delta E_{N,N+1}$.